# IEEE 802.15.4.e TSCH-Based Scheduling for Throughput Optimization: A Combinatorial Multi-Armed Bandit Approach

Nastooh Taheri Javan, Masoud Sabaei* and Vesal Hakami[1]

*Abstract*—In TSCH, which is a MAC mechanism set of the IEEE 802.15.4e amendment, calculation, construction, and maintenance of the packet transmission schedules are not defined. Moreover, to ensure optimal throughput, most of the existing scheduling methods are based on the assumption that instantaneous and accurate Channel State Information (CSI) is available. However, due to the inevitable errors in the channel estimation process, this assumption cannot be materialized in many practical scenarios. In this paper, we propose two alternative and realistic approaches. In our first approach, we assume that only the statistical knowledge of CSI is available a priori. Armed with this knowledge, the average packet rate on each link is computed and then, using the results, the throughput-optimal schedule for the assignment of (slot-frame) cells to links can be formulated as a max-weight bipartite matching problem, which can be solved efficiently using the well-known Hungarian algorithm. In the second approach, we assume that no CSI knowledge (even statistical) is available at the design stage. For this zero-knowledge setting, we introduce a machine learning-based algorithm by formally modeling the scheduling problem in terms of a combinatorial multi-armed bandit (CMAB) process. Our CMAB-based scheme is widely applicable to many real operational environments, thanks to its reduced reliance on design-time knowledge. Simulation results show that the average throughput obtained by the statistical CSI-based method is within the margin of 15% from the theoretical upper bound associated with perfect instantaneous CSI. The aforesaid margin is around 18% for our learning-theoretic solution.

## I. INTRODUCTION

The Internet of Things (IoT) has been steadily emerging in a wide variety of applications in recent years [1]. One of the promising applications of IoT is the Industrial Internet of Things (IIoT). IIoT connects a large number of industrial devices to the Internet in order to implement new applications such as factory automation, distributed process control and real-time monitoring [2]. One of the platforms that the IIoT applications are being implemented on today is the Industrial Wireless Sensor Networks (IWSNs), whose requirements include low latency, robustness and determinism [3].

Many WSN implementations in the physical layer and MAC layer adopt the IEEE 802.15.4 standard. This standard, despite its advantages and popularity, is not able to meet the requirements of IWSN's industrial applications in terms of reliability, transmission rates, latency and energy consumption [4]. To overcome these weaknesses, as well as to define a low-power multi-hop MAC protocol, IEEE has established a working group, called IEEE 802.15.4e, to add these capabilities to existing 802.15.4. The IEEE 802.15.4e amendment provides three MAC mechanisms that are optimized for different automation domains, such as the Low Latency Deterministic Network (LLDN), the Deterministic Synchronous Multi-Channel Extension (DSME) and the Time Slotted Channel Hopping (TSCH). LLDN has been designed for very low-latency applications, such as robotics and car manufacturing in single hop and single channel networks. DSME targets industrial, commercial and health applications, such as telemedicine and smart metering, and TSCH is designed for application domains such as process automation in multi-hop and multi-channel networks [4].

TSCH combines time slotted access with multiple channels and Channel Hoping. This offers benefits that includes higher network capacity, greater reliability, predictable latency and reduced energy consumption [5]. The main communication unit in TSCH is the slot-frame, which is required by a pair of nodes in the network to exchange data. Slot-frame is a set of timeslots that is repeated continuously over time. The TSCH protocol uses a slot-frame to provide a synchronized connection in the network. For each timeslot, a different channel is assigned pseudo-randomly and the schedule tells which neighbor to communicate with and on which channel offset.

In the TSCH protocol, only the execution time of the MAC scheduling is specified without dictating the scheduler's execution manner [6]. In fact, this protocol does not specify how to create and maintain a proper link scheduling. In practice, scheduling in TSCH specifies the frequency and slots for each link of a node. In recent years, many researchers have addressed the scheduling problem for the TSCH protocol, from centralized to distributed solutions [6].

Nastooh Taheri Javan and Masoud Sabaei are with the Computer Engineering department of Amirkabir University of Technology (Tehran polytechnic), Tehran, IRAN. (e-mails: nastooh@aut.ac.ir, sabaei@aut.ac.ir).

Vesal Hakami is with the School of Computer Engineering, Iran University of Science and Technology, Tehran, IRAN. (e-mail: vhakami@iust.ac.ir).
*Masoud Sabaei is the corresponding author.

However, channel-aware scheduling has not received much attention in previous studies. In fact, the majority of the existing solutions have been mainly concerned with autonomous, distributed assignment of slot-frame cells to communication links, see [7-10]. These schemes deal with the distributed computation of local schedules with varying objectives, such as reduced signaling overhead, minimal duty cycle (reduced idle listening), improved energy efficiency, reduced interference, and minimal number of conflicts between the nodes. There are also sundry other publications, [11-14], which address data queue-aware TSCH-based scheduling and investigate the impact of different traffic types without explicitly considering the impact of link qualities or channel variations.

A small number of related papers tackle channel-aware throughput optimization in their formation of a TSCH schedule. As these schemes are more closely related to our proposed scheme, we elaborate on their underlying assumption regarding the availability of CSI: First, there are those, see [15, 16 and 17], that compute a throughput-optimal schedule by relying on the unrealistic assumption that a perfect non-causal knowledge of instantaneous channel qualities (e.g., instantaneous SNR) is available for an arbitrary-length time horizon. Second, there are other scheduling algorithms [18, 19 and 20] that utilize prior statistical knowledge of link qualities (e.g., packet error rate (PER) or expected number of transmissions (ETX)) to improve average packet delivery ratio (PDR). A more relaxed assumption on CSI availability has been made in [21], where a wavelet-based estimation of link qualities is presented for the real-time computation of TSCH schedules. However, the procedure described in [21] is still dependent on some prior probabilistic model of the noise in the wireless environment.

In wireless communication networks, CSI is influenced by several factors in a random and time variant manner, such as signal scattering, fading, channel gain, and power loss ratio with the distance between the transmitter and the receiver [22]. In these networks, the instantaneous CSI estimation is one of the most challenging issues. In particular, the usual way to estimate CSI is to send a pilot signal and receive its feedback. In this manner, CSI on the sender's side is not fully accessible due to faulty or delayed feedback and frequency offset between the mutual channels. Moreover, in fast-fading networks, instantaneous CSI loses its credibility quickly, so it is not reliable. Consequently, in practical systems, it is impossible to obtain the instantaneous CSI in advance [23]. As such, the basic assumption in earlier scheduling algorithms, where channel status is available or constant, is not realistic.

In this paper, we explore two centralized approaches to TSCH scheduling that deal with the issue of CSI availability: In our first approach, the assignment of slot-frames is based on the statistical knowledge of the channel (rather than on the exact instantaneous knowledge), which is assumed to be a priori known at design time. Then, an optimal scheduling in terms of the average throughput of the entire network is computed. In fact, under this assumption, the average number of packets sent by each connection in each slot can be obtained in advance, and then the timing is optimized based on the average connection rates. In this approach, using the famous Hungarian algorithm [24] in graph theory, we compute the max-weight assignment in polynomial time. It is important to note, however, that in real circumstances, the statistical information on CSI either changes over time or is not available a priori, therefore, we propose our second approach.

In our second proposed scheme, we avoid reliance on the existence of CSI and knowledge of its probabilistic model, hence it can be used in a wider range of operating environments. Moreover, we apply a machine learning-based method to compensate for the lack of design-time knowledge. Here, instead of relying on a predetermined model, we draw on the experience gained from real interaction with the operating environment to create the scheduling. This second approach is intended to be our main contribution in this study.

In particular, we recast the scheduling problem as a combinatorial multi-armed bandit (CMAB) process [25], which is an extension of the classical multi-armed bandit (MAB) framework in machine learning theory [26]. The MAB framework is used to address the exploration-exploitation dilemma faced by a learning agent operating in an unknown uncertain environment. In the classical MAB problem, action choices are scalar (one dimensional), and the agent must decide which arm to play at each round so as to maximize its average reward over the long run. In CMAB, however, the agent selects not just one arm in each round, but also a subset of arms or a combinatorial object in general, referred to as a super arm. The reason why we model the problem as CMAB rather than as MAB, is because of the nature of the scheduling process in multi-channel wireless networks with channel hopping. The scheduling problem in TSCH is actually a matching (assignment) problem in which a set of arms are played together in each round. Based on our CMAB formulation, we then develop a scheduling algorithm which is inspired from the Linear Learning Rewards (LLR), see [27], as a general online solution for MAB problems in combinatorial settings.

The contributions of this paper are as follows:

- A TSCH-based scheduling algorithm is derived for throughput optimization in IEEE 802.15.4e networks by considering statistical CSI instead of instantaneous CSI. This is done by modeling a bipartite graph with all subset of non-interfering links as vertices of the upper side, and slot-frame matrix cells as the lower set of vertices of the graph. By knowing the average packet rate per connection, optimal assignments are calculated using the well-known Hungarian algorithm [24].

- Our second solution, as our main contribution, addresses those settings where not even the statistical CSI is available at design time. In such zero-knowledge settings, a machine learning-based approach (using CMAB formalism) is proposed to compute the optimal schedule based on real-time interactions with the wireless network.

- Aside from its practical value for IEEE 802.15.4e settings, the problem addressed in this paper also showcases an interesting application of CMAB to networking problems.

The rest of the paper is organized as follows: in Section II, previous studies on scheduling for TSCH networks are investigated. In Section III, our system model and the assumptions are described. In Section IV, the problem is formulated on the basis of statistical and unknown CSI and the proposed solutions are presented. In Section V, our simulation results are presented, and finally Section VI concludes the paper.

## II. BACKGROUND

In the following section, the TSCH scheduling protocol is briefly reviewed, and then a number of scheduling algorithms are investigated.

### A. Scheduling in TSCH

In TSCH networks, scheduling (i.e. assigning links to nodes to send data) helps to effectively allocate wireless links in order to maximize the number of communications. The specifics of scheduling have a direct impact on network efficiency, including throughput, node energy consumption, flow latencies, reliability, and overhead. A scheduling pattern defines and specifies the time slot and channel in which each node should send data to or receive data from its neighbors. The main unit of the scheduling bandwidth is called a cell. The length of each cell today is typically 10 ms, during which the transmitter sends the data packet and the receiver returns the corresponding acknowledgment following a successful reception.

In the TSCH, a channel hopping pattern containing 16 offsets for sending on multiple frequencies is defined by default. It should be noted that in each time slot a node changes the physical channel pseudo-randomly by combining the channel offset and the ASN (Absolute Slot Number). In particular, the frequency $f$ can be derived as follows:

$$f = Fnc\{(ASN + chOffset)\ mod\ nChannels\} \quad (1)$$

where $chOffset$ indicates the channel offset, $nChannel$ denotes the number of available physical channels and $Fnc$ is the mapping function.

### B. Related Works

From one perspective, TSCH scheduling schemes can be divided into two groups: centralized and distributed. In centralized schemes, a central entity collects the entire network information and then calculates the link schedules. In a distributed scheme, the neighboring nodes exchange their scheduling information, and each node determines its scheduling based on local information. In a centralized approach, while the scheme provides integrated vision and scheduling without collision that results in improved reliability, it suffers some weaknesses such as a time consuming initial setup, lack of flexibility when faced with sudden network changes, signaling overhead in terms of number of messages and higher energy consumption to set up and maintain scheduling. In contrast, distributed approaches are considered more suitable for large-scale networks because they are usually responsive to sudden network changes. Of course, these schemes also have their own demerits such as the need for negotiation between neighbors.

In almost all previous TSCH scheduling, authors either assume that the impact of link qualities or channel variations on performance are negligible (such as [7-14]), or that the instantaneous and complete channel information is available (such as [15-17]), or else the channel is modeled quasi-static (such as [18-20]). For example, in [15 and 16], it is assumed that the instantaneous CSI is fully measurable and available, therefore one can use an offline method and calculate the optimal scheduling for a TSCH network by the exact rate of packets to be sent over each link. In particular, in [15] authors present a graph and matching theory-based approach to maximize throughput. To solve this problem, they applied the Hungarian algorithm [24], and came up with a solution that generated optimal throughput in the presence of channel information. In [16], TSCH scheduling is modeled with the goal of energy efficiency maximization, and an optimal greedy method based on the Vogel's approximation method [28] is presented. In this method, among all the nodes that apply for a specific channel, a node with the highest amount of remaining energy is selected. Simulation results show that the proposed approach consumes less energy than many of the investigated methods.

In [18], a $k$-cast scheduling is introduced for TSCH networks, where $k$ different receivers are allocated to the same transmitter in order to increase the probability that at least one device receives the packet correctly. In this method, the reliability is improved by using a multi-path routing approach. Authors investigated the impact of their proposed approach on delay, jitter, network capacity, energy consumption and delivery ratio. In [20], the ReSF technique is presented as a scheduler for TSCH networks. It tries to compute paths with the least delay from source nodes to sink. In this approach, authors modeled the scheduling problem as an Integer Liner Program (ILP) to provide a way to measure the packet loss rate, collision rate, and latency. It is clear from their simulation results that in this case superior performance on delay is achieved compared to the investigated solutions. In [19], a distributed scheduling method with local blacklist configuration is proposed for TSCH networks, called LOST. Authors investigated the impact of cell over-provisioning on transmitting a single packet over various link qualities, to identify the minimum required number of cells.

In [11], authors present AMUS, which is a focused and traffic-aware scheduler for real-time industrial applications. In this approach, in order to reduce idle listening, more resources are allocated to weak and vulnerable links related to relay nodes near the sink. In [12], CLS, a centralized multicast scheduling, is introduced. It focuses on the reduction of idle listening, while addressing bandwidth requirements of the network. The basic idea behind CLS scheduling is that instead of generating the entire schedule every time, the sink node reduces the signaling overhead of the algorithm only by assigning and retaining some time slots. Simulation results show that in this signaling scheme overhead and idle listening are lower than those in some of the competing algorithms.

Palattella et al. [13] have proposed a centralized scheduling called TASA which is an attempt at reducing delay in TSCH networks. The TASA algorithm requires a central node as a coordinator to know the full network topology and traffic load that each node produces in each slot. In [29], authors presented a distributed version of TASA, called DeTAS, to create a collision-free optimized multi-hop scheduler. The goal was to minimize buffer overflow and several coordinators were used in this method to achieve this goal.

In [7], Wave, a distributed scheduling algorithm is presented. The main purpose of Wave is to try to create a schedule in which the number of allocated slots is minimized while the throughput is increased and the delay is decreased. Authors concluded that the slot-frame size can be reduced by dividing it into units, called wave, to create scheduling for all nodes. In [10], a fully distributed scheduler named ASS (Adaptive Static

Scheduling) is presented. In ASS, each pair of nodes activates a set of timeslots in order to improve the energy efficiency dynamically. In this method, the nodes can increase or decrease the allocated timeslots according to the traffic pattern. However, the study did not explore the cost of blind over-provisioning in terms of delay performance.

A new distributed scheduling scheme for TSCH networks, titled DeAMON, is presented in [9] for industrial monitoring and control applications. Goals considered for the DeAMON scheduler include traffic-aware scheduling, simultaneous transmission support, on-demand topology changeability, and signaling overhead reduction. The solutions offered by this scheduler are for upward traffic only and do not pay attention to downward traffic. Also, in [14], authors introduced a new distributed scheduling, called DIVA. In this approach a network node can be in one of three states: sending, receiving or idle. At first, the sender node sends a connection request to the receiver on the signaling channel, then the receiver returns an ACK to the sender in order to establish a new connection. Subsequently, the sender node changes its state to sending and consequently the receiver node changes its state to receiving and selects a channel for communication, randomly.

In [8], a scheduler named Orchestra is introduced. Orchestra is a no-graph autonomous scheduling in which each node calculates its own specific scheduling based on the RPL [30] routing specification. In this scheduler, there is neither central control node nor signaling. For these reasons, this scheme can be considered a simple and flexible scheduler and generally different from other timing algorithms for TSCH networks. In Orchestra, each node anonymously calculates and maintains its schedule independent of signaling, and updates it automatically, without any signaling overhead. The Orchestra scheduler is not suitable for scenarios where different nodes require different bandwidths with their neighbors.

In [31], a blacklisting-based link assignment approach is developed in which every two nodes in the network creates a blacklist of channels between them locally. A number of channels are assigned to each time slot, and each link in the cell uses non-black-listed channels. The quality of the channel in the physical layer is estimated by Multi Armed Bandit optimization (MAB).

### C. Motivation

In channel-aware scheduling, CSI can be available either as non-causal CSI, which contains CSI of the past, present, and future slots, or causal CSI, which contains only CSI of the past and present slots. To the best of our knowledge, there are many studies which present new scheduling methods for TSCH networks without considering the impact of CSI (such as [7-14]). We have seen that in some schemes authors assume that perfect non-causal knowledge of CSI is available for the scheduling process (such as [15-17]) and in others prior statistical knowledge of link qualities is utilized in order to improve performance (Such as [18-20]). Although such assumptions in earlier scheduling algorithms, where channel status is available or constant, they provide mathematical tractability of the scheduling problem, they are not realizable in real-world systems due to the difficulty in predicting random and time-varying wireless channel conditions. In this paper we seek to relax the CSI availability assumption for TSCH-based scheduling.

## III. SYSTEM MODEL AND ASSUMPTIONS

### A. Network Model

We consider a TSCH network consisting of $N$ nodes including a gateway such as in Figure 1. The network nodes are managed by the gateway, in which resides a scheduler that determines for each node how many packets and in what channel should be transmitted during a time slot

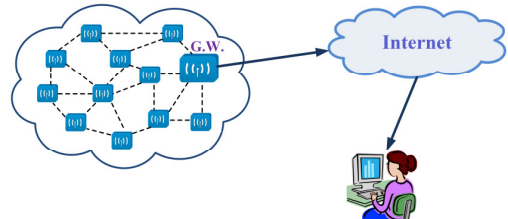

Fig. 1. An IoT network configuration.

The physical topology of the network is modeled as a graph $G = (V, E)$ in which $V = \{n_0, n_1, \ldots, n_{N-1}\}$ is the set of nodes and $E$ is the set of links between the two nodes $e_{ij} = (n_i, n_j) \in E \subset V \times V$ in the network. Each node $n_i$ is equipped with a radio, having a communication range of $R_i$ that is potentially larger than the interference range $\hat{R}_\iota$. The symbols $t \in \{1, \ldots, T\}$ and $f \in \{1, \ldots, F\}$ denote the moment of each slot and the set of frequencies in the network, respectively, and $\varDelta$ is the duration of each slot.

Figure 2 illustrates an example of a slot-channel matrix in a TSCH network for a 4 node configuration graph $G$. Some cells are dedicated, while the others are shared between multiple links, such as $B \rightarrow A$ and $C \rightarrow D$. Here, a slot frame with 3 slots and 4 channels is considered. Each connection is a transaction that occurs in a cell of the slot-frame.

It is assumed that the nodes have only one half-duplex radio module and that they are able to use the radio module at different times for sending or receiving on different channels, but restricted to send or receive on a single channel.

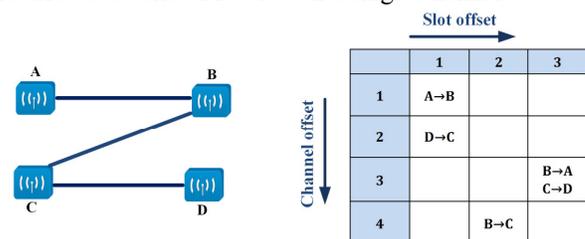

Fig. 2. Slot-channel matrix for a TSCH network example

### B. Traffic Model

We assume that all nodes have saturated traffic model in the sense that traffic is always backlogged at each node. A saturated source has elastic traffic so that it appreciates as much transmission rate as it gets without concern for a strict deadline requirement. To operate efficiently under this traffic type, we need throughput optimization which only needs channel-awareness, a thing we consider extensively in this paper.

***Remark 1:*** The saturated traffic assumption has been conveniently made to justify our uni-criterion throughput-centric analysis. We are aware that in more realistic dynamic traffic scenarios, depending on the traffic load at each node, a node may or may not have data to transmit when the medium has become idle. In the sequel (c.f., Remark 2), we elaborate more on the technical difficulties associated with relaxing this assumption, and still maintaining the model-free nature of a TSCH scheduler.

It should be noted, however, that although the traffic model is assumed to be saturated and non-stochastic, but still the channel condition varies randomly with time. Hence, we need to base our design on average throughput or the average transmission rate that is achievable using a fixed power level over a link with stochastic channel quality. In case the statistical knowledge regarding the channel condition is available, one can compute the expected number of packets that can be transmitted over each link.

### C. Channel Model

The state of each channel $f \in \{1, \dots, F\}$ during slot $t$ is shown by $X_{f,t}$. It is assumed that $X_{f,t}$ is constant over slot $t$, and only changes at the slot boundaries. If the user sends a signal $Y_t$ at the $t$-th slot, the signal received $S_t$ is in accordance with (2):

$$S_t = H_{f,t} Y_t + W_t \tag{2}$$

where $H_{f,t}$ indicates the gain of the channel $f$ in relation to the fading phenomenon and $W_t$ represents the zero mean noise with the variance $N_0^2$. Also, $X_{f,t} = |H_{f,t}|^2$ is the channel state at time $t$. The power required for reliable and error-free communications in the case where $X_{f,t} = x$ and the transmitted signal corresponds to transmitting $u$ packets, each of $l$ bits, is in the form of equation (3), thus:

$$P(x, u) = \frac{\beta N_0}{x} 2^{\frac{ul}{\beta}} - 1 \tag{3}$$

where $\beta$ indicates the bandwidth of the received signal.

The transmission power is assumed to be constant, so following the discussion in [32], the total number of packets that can be sent over the link $e$ when its channel state is at $X_{f,t} = x$, can be obtained from equation (4) as follows:

$$U_e(x) = \frac{\beta}{l} \log \left( 1 + \frac{xP}{\beta N_0} \right) \tag{4}$$

Although the gain of the channel $H_{f,t}$, hence $X_{f,t}$, may be a continuous random variable, it is assumed for simplicity that $X_{f,t}$ only takes values from a finite state space $\psi$. Actually, a quantized model is assumed for the channel random status. The channel status is assumed to be an i.i.d random process, which takes values from $\gamma$ different levels with the boundaries specified in (5):

$$\{(-\infty, \Gamma_1), [\Gamma_1, \Gamma_2), \dots, [\Gamma \gamma, \infty)\} \tag{5}$$

We can make such a division for all channels, provided the probability distribution of the channel status at each of these levels is different for each channel and each link $e_{ij} = (n_i, \ n_j) \in E$ (as discussed in [33]).

### D. Interference Model

It is assumed that interference will occur if two nodes $(n_i, n_j) \in E$ send data simultaneously to the same receiver. The transmission success condition of the node $n_i$, as shown in Figure 4 is:

1)   $d_{ij} \leq R_i$

2)   For each node $n_k \cdot d_{kj} \leq \hat{R}_k$

As shown in Figure 3, the node $i$ lies within the range $R_i$, and the node $k$ is also within the range $R_k$. The node $j$ is located in the radio range of both $i$ and $k$. Collision will occur in case of a simultaneous transmission to the node $j$. In other words, a node cannot send and receive at the same time, nor can it simultaneously receive from multiple nodes.

We define a collision graph $Q = (E, C)$ to take into account the interference in the problem formulation. Its vertices correspond to the edges of the configuration graph $G$, and its edges indicate the interference between the two links. Figure 4 illustrates the different modes of sending data in a collision graph $Q$. The transmissions are unicast, hence transmissions such as $B \rightarrow C$ and $B \rightarrow A$ can not take place simultaneously, and there is an edge between them in the collision graph. Therefore, a valid schedule causes this data not to be sent in the form of a shared cell. Also, there will not be any edge in the collision graph from a vertex to itself.

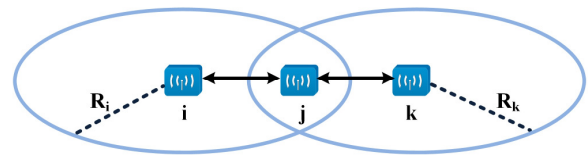

Fig. 3. Successful Transmission of a Node

In fact, a scheduling algorithm should select a so-called "independent set" of the vertices in the collision graph to be scheduled in the same cell. Recall that an independent set in a graph is a set of vertices such that there are no edges between each two vertices. The goal of scheduling is that two interfering nodes are not scheduled for transmission in the same physical channel.

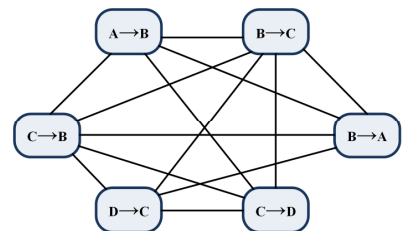

Fig. 4. Collision Graph of Figure 2.

In this model, the links within a given independent set in the collision graph can be scheduled simultaneously using the same physical channel (i.e., within the same slot-frame cell). This computation is assumed to be done in a pre-processing step before running the actual algorithms for computing the schedules. Now, due to the self-reducibility of the independent set problem, fast and efficient algorithms exist that can list all independent sets of a general graph [34]. These algorithms have low-degree polynomial worst-case time and memory footprint between any two consecutively output solutions. More discussion regarding the complexity of this pre-processing step is given at the end of Section IV.A.

## IV. Problem Definition and Proposed Solution

In this section, we model and formulate scheduling of slot-frames for a TSCH network as a problem of "maximum-weighted matching in a bipartite graph", similar to the approach in [15]. Each subset of the non-interference links between the IoT nodes is modeled as a vertex of the upper side of the bipartite graph and each slot-frame matrix cell is also considered to be the bottom vertices of the graph. The weight of each edge is equal to the total number of packets to be sent in case where assigning the desired cell from the slot-frame matrix to the corresponding link. In this way, a valid (non-interference) scheduling would be equivalent to a "full assignment" in this bipartite graph. Since our goal is maximization of the network throughput, we seek an assignment that generates the maximum total weight of the edges.

The method introduced in [15] relies on the assumption of the availability of CSI to calculate the best match, which is not realistic in practical scenarios, although it has less signaling overhead compared to our proposed method. In [15], it is assumed that the instant CSI is measurable in its entirety. Thus, it is possible to calculate the optimal scheduling using an offline method by having the exact rate of packets to be sent on each link. In wireless communication networks, in a random and time variant manner, CSI is influenced by several factors such as signal scattering, fading, channel gain, and power loss ratio with the distance between the transmitter and the receiver [22], hence, estimating instantaneous CSI is very challenging in telecommunication. Usually, CSI is estimated by sending a pilot signal and receiving the feedback. But due to erroneous or delayed feedback and frequency offset between the mutual channels, CSI is not fully accessible. Additionally, instantaneous CSI tend to be invalidated quickly in fast fading networks.

Given the above argument, and in contrast to the instantaneous CSI approach adopted in [15], we propose the following two solutions:

1. We first assume that at the optimization time, only statistical CSI (i.e. the probability distribution of channel fading) is available. Compared to instantaneous CSI, reliance on statistical CSI leads to a more desirable solution, especially in scenarios with faster channel variation rates. Knowing the link quality probability distributions, we can calculate the average link packet rates and consider them as the weight of the bipartite graph edges. Given the edge weights, the maximal weight assignment calculation is carried out within a polynomial time using the famous Hungarian algorithm [24]. However, in many scenarios, it is not possible to accurately model a CSI probabilistic structure. In addition, the solution for a specific model loses its validity with changing circumstances, and no longer applies to other operational environments, so we move on to our second approach which addresses the problem from a different perspective and works free of CSI.

2. In this approach, we assume that there is no information on the random status of the channel at the decision making instance, either as an instantaneous measurement, or in the form of a pre-determined statistical distribution. We make the assumption that, after applying scheduling decisions to the network, the TSCH scheduler can receive feedback from nodes (for example in the form of per link packet rate report.) By receiving feedback, according to the feedback history, the scheduler will update its next decision-making policy and re-apply it to the network. In this way, the second proposed solution is a kind of machine learning process that should be able to solve the weighted maximum assignment problem in real time by interacting with the environment, gaining experience and sampling the random system dynamics. This solution is of particular importance not just due to its model-free nature, but also because it relies neither on erroneous CSI measurements nor on a pre-built statistical model of link status.

In our next section, the problem is formulated once by assuming a statistical CSI and then by assuming no CSI. This is followed by presentation of the proposed algorithms.

### A. Problem Formulation by Assuming Statistical CSI

In this section, we model the throughput maximization problem (under the protocol model) in the case of known channel random distribution as a prelude to our second approach. Our goal is to construct a list of the frequencies and time slots for the links $e \in E$, which maximizes the total network throughput by sending packets based on that list.

Denote by $\bar{U}_e$ the average number of packets that can be sent over link $e$ (in case of assigning a cell from the slot-frame to it), as defined by equation (6):

$$\bar{U}_e = \mathbb{E}[U_e] = \frac{1}{F} \sum_F \sum_x \mathcal{P}(x) U_e(x) \tag{6}$$

If we define $\mathbb{I}_{e,f,t}$ as a binary decision variable, the mean throughput maximization problem can be formulated as follows:

$$U^* = \max \sum_{e \in E} \sum_{f=1}^{F} \sum_{t=1}^{T} \mathbb{I}_{e,f,t} \, \bar{U}_e \tag{7}$$

s.t.

$$\sum_{f=1}^{F} \sum_{t=1}^{T} \mathbb{I}_{e,f,t} \geq 1; \ \forall \, e \in E, \ \forall \, f \in F, \ \forall \, t \in T \tag{8}$$

$$\mathbb{I}_{e,f,t} + \mathbb{I}_{\acute{e},f,t} \leq 1; \ \forall \, e, \ \acute{e} \in N, \ e \neq \acute{e}, \ \forall f, \forall t \tag{9}$$

$$\mathbb{I}_{e,f,t} \in \{0,1\} \, ; \ \forall \, e \in E, \ \forall \, f \in F, \forall \, t \in T \tag{10}$$

Equation (7) shows our objective function as the maximum expected total throughput of all nodes in the TSCH network in which the constraint in equation (8) ensures that each node is assigned to a time slot, thus meeting the requirement for fairness. This relationship is not applicable in all scenarios because there may not be enough slots in a slot frame. In other words, if more bandwidth is required for link $e$, by placing more slots for that link, it is possible to send more packets in a frame. The constraint in equation (9) is to prevent collisions, which guarantees that a maximum of one user can transfer data in a specific slot and channel offset.

According to the above formulation, we can model the problem of total network throughput maximization as a weighted optimal assignment problem in a bipartite graph. In this graph, the vertices are divided into two disjoint sets (the upper set and the lower set). Each edge joins a vertex from one set to a vertex from the other set. A valid assignment is a set of non-connected edges, i.e., no two edges share a common vertex. A perfect assignment is one in which each upper vertex in the bipartite graph corresponds to exactly one vertex of the lower set of the graph (see Figure 5).

Suppose $\hat{G} = (\acute{U}, \acute{V}, \acute{E})$ is a weighted bipartite graph, which, in accordance with Figure 5, the set of vertices in the bottom is denoted by $\acute{U} = \{\acute{u}_1, \acute{u}_2, ..., \acute{u}_N\}$, and corresponds to the slot-frame cells. Also the set $\acute{V} = \{\acute{v}_1, \acute{v}_2, ..., \acute{v}_f\}$ denotes the set of vertices at the top, and corresponds to the collision graph nodes.

The set $\acute{E} = \{\acute{e} = (\acute{u}, \acute{v}) | \acute{u} \in \acute{U}, \acute{v} \in \acute{V}\}$ are the edges of the graph $\acute{G}$. The weight of edge $\acute{e}_{pq} = (\acute{u}_p, \acute{v}_q)$ is equal to $w_{pq} \geq 0$, which can be calculated from equation (6).

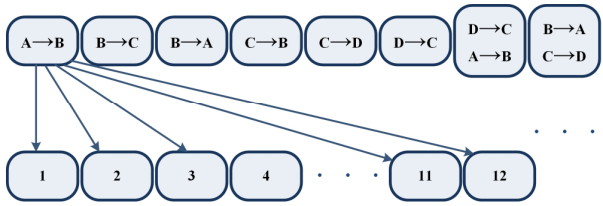

Fig. 5. Corresponding Bipartite Graph

One subtlety in the definition of weights is that in our bipartite graph, the weight of each link from a node $\acute{u} \in \acute{U}$ to all the nodes $\acute{v} \in \acute{V}$ is identical. This is due to the basic channel hopping process in default TSCH. In fact, over the course of one slot-frame cycle, each slot-frame cell cycles through all physical channels. Hence, in the "average" sense, and for a given particular communication link, there is no difference between one cell or another, as all cells will eventually experience all physical channels over time (and in a uniform pseudorandom fashion). Therefore, when computing the weights on link-to-cell mappings, we need to consider identical weights for all the edges ($\acute{u}, \acute{v}$), $\forall \acute{v}$ corresponding to the mean throughput (across all frequencies) to capture the impact of TSCH frequency hopping. Obviously, these weights might vary for another $\hat{u} \in \acute{U}$ as different communication links are associated with different channel quality distribution.

To guarantee feasibility (i.e., the existence of perfect assignment), we assume that we can also have zero-weight edges. This assumption is also in line with the basic idea behind IEEE 802.15.4e TSCH in that if there is nothing to send on each node, it always sends out the idle packets. Sending an idle packet represents the zero weight for the corresponding edge.

Now, based on this bipartite graph, we may use the Hungarian algorithm [24] in graph theory to determine the optimal weighted assignment. The Hungarian algorithm is a low-complexity algorithm which operates in (3rd degree) polynomial time in terms of the number of vertices of the input bipartite graph. In our case, the number of vertices in graph $\acute{G}$ corresponds to the number of independent sets in the collision graph $Q$. In general, finding an upper bound for the number of independent sets in different types of graphs is a problem in its own right (e.g., see [35]). Pessimistically, the trivial upper bound of $2^{|E|}$ (for a topology graph with $|E|$ links) is as tight as one can get, since the extremely sparse collision graph that has no edges does indeed have $2^{|E|}$ independent sets. However, we argue that in most practical settings, the running time associated with our matching problem will not be prohibitively high, especially in moderate scales and practically dense topologies.

### B. Problem Formulation under Unknown Channel Conditions

Now, in contrast to the formulation described in Section IV.A, in this section, we assume that the statistical distribution of channel random status, $X_{f,t}$, is not available. We aim to solve the same problem of optimal assignment with random weights and unknown distribution in an online manner. To formulate the problem under unknown CSI conditions, we use the Combinatorial Multi-Armed Bandit (CMAB) [25] framework.

The CMAB framework is, in fact, an extension to the classical and simpler problem of Multi-Arm Bandit (MAB) [26]. Now, before addressing the formulation of the problem, we briefly provide a background to the MAB problem.

#### 1) MAB Problem in Learning Theory

MAB is a classic formulation framework in learning theory, in which an agent must choose between several arms that each time yields some random reward. The purpose of the agent is to identify the arm with the highest average reward. If the random distribution of rewards for each arm is already known, the agent could (prior to any selection) identify the arm with the highest average reward and receive the current rewards from this arm. But the main challenge is to identify the arm with the highest average reward without having a random system model.

In order to solve the learning problem, the MAB agent must, while making a decision, establish a balance between two types of exploration and exploitation decisions. On the one hand, all the arms should be selected in order to test and learn enough to estimate their average weight. On the other hand, the previous observations should be exploited by selecting the greedy arm of the best-rated weight for the highest average reward. Incidentally, the expansion of MAB to CMAB can be used in combinatorial problems for a much wider range of optimization problems. The main challenge in CMAB is the exponential number of the arms to be selected by the agent. For example, in the assignment problem, the total number of complete assignments in a bipartite graph with $n$ vertices on one side is equal to: $n!$. For this reason, the classic MAB learning algorithms will not be applicable for CMAB.

In this paper, we use the LLR learning method for CMAB problems, presented in [27], to learn the optimal assignment under unknown CSI conditions. The main idea of this algorithm is to exploit the evolutionary property of any vertex of the bipartite graph in different arms to meet the need for frequent and separate sampling of each arm (the full assignment of our problem). The average weight of each edge is estimated over time and updated for each feedback received from each new assignment. All the edges are adequately sampled and examined over time. After sufficient observation, the weight of the entire edges is estimated with an appropriate approximation method and, consequently, full optimal assignment is also learned.

#### 2) CMAB-Based Formulation

We consider the TSCH scheduler for assigning slots frames to network links. We show the weights of each edge $1 \leq e \leq N$ at the instant $t$ with $U_e(t)$. Given the random nature of CSI, $U_e(t)$ is an i.i.d. random process over time. Its values can be normalized in the interval $U_e(t) \in [0, 1]$. Contrary to section IV-A, here it is assumed that due to lack of statistical knowledge of the channel state, the mean $\overline{U}_e = \mathbb{E}[U_e]$ is unknown.

Let's consider a slot frame cycle with index $\tau$ during which an assigned link to a cell jumps on all $F$ frequency offsets. During the cycle $\tau$, we show a decision for scheduling with the decision vector $a(\tau)$ having the dimension $\acute{E}$, which selects a complete assignment set. In fact, for $1 \leq \acute{e} \leq \acute{E}$ we have $a_{\acute{e}}(\tau) \geq 0$. When a specific $a(\tau)$ is determined, in each time slot $n$, only for each $\acute{e}$ with $a_{\acute{e}}(\tau) \neq 0$, the feedback value of $U_{\acute{e}}(X_{f,t})$ is reported by the sender node(s) to the TSCH scheduler.

In fact, when a transmission is done on the quality status $X_{f,t}$ of the channel $f$, $U_{\acute{e}}(X_{f,t})$ represents the number of packets sent during the time slot $n$ from the cycle $\tau$ on the links belonging to the upper vertex of the $\acute{e}$ in the bipartite assignment graph. This value is calculated according to equation (6). Hence, the reward obtained during cycle $\tau$ is given as (11):

$$U_{a(\tau)} = \frac{1}{F}\sum_{f\in F}\sum_{\acute{e}\in\acute{E}} a_{\acute{e}}(\tau)U_{\acute{e}}(X_{f,t}) \qquad (11)$$

In (11), using the decision vector $a_{\acute{e}}(\tau)$, a set of complete assignments (i.e., a "perfect matching" in graph theory terms) is selected, and its overall weight is calculated as the total number of packets sent over time slot $t$. To capture the impact of the default frequency hopping in TSCH (c.f., Section II.A) , the same action $a(\tau)$ will be carried out over the entire cycle $\tau$, and the mean throughout (across all frequencies) is recorded as the overall reward obtained during cycle $\tau$.

In the framework of CMAB, we evaluate the performance of the learning algorithm with respect to regret[2] [25]. The regret is calculated as the difference between the average realized reward during the learning process and the expectation of the optimal reward. Minimizing regret is equivalent to maximizing reward. In the cycle $\tau$, regret is defined by equation (12):

$$Z(T) = tU^* - \mathbb{E}[\sum_{\tau=1}^{T} U_{a(\tau)}] \qquad (12)$$

We want to reduce $Z(T)$ as much as possible when $T \to \infty$. As a result, the maximum reward is obtained on average.

In the next section, using the results presented in reference [27], we propose an algorithm that has an $O(N)$ storage memory and is able to bring the regret close to zero over time.

### 3) Proposed Algorithm for Calculating Optimal Assignment

Algorithm 1 shows pseudo-code of the proposed algorithm. Lines 2 to 7 correspond to the initialization phase of the algorithm. During this phase, the algorithm makes sure to have an initial estimate for the weight of each edge in the bipartite scheduling graph. Hence, in line 5, it loops through every edge of the graph, and using the standard Hungarian algorithm [24], finds a perfect assignment $a(\tau)$ within which the current edge is necessarily included. It then calls the Evaluate() method (line 6) to update two parameters $\theta_{\acute{e}}$ and $m_{\acute{e}}$, where $\theta_{\acute{e}}$ indicates the current estimate of the mean weight of link $\acute{e}$, and $m_{\acute{e}}$ tracks the number of times link $\acute{e}$ has been visited within any given matching from the very beginning of the execution. $\theta_{\acute{e}}$ is updated using a simple moving average process as given in (13), and $m_{\acute{e}}$ is a just a counter updated according to (14).

Our algorithm time corresponds to slot-frame cycles, indexed by $\tau$. Hence, in the Evaluate() method, the current schedule $a(\tau)$ is applied for an entire cycle to the network so as to capture the impact of the default TSCH mechanism for cyclic frequency hopping. At the end of cycle $\tau$, $U_{\acute{e}}(\tau)$ gives the latest observation from the weight over the edge $\acute{e}$, and is then used in the update equation for $\theta_{\acute{e}}$ to obtain a recent estimate of the mean weight of $\acute{e}$.

The second phase of the algorithm (i.e., MAIN LOOP) is basically similar in spirit to the INITIALIZATION phase; however, this is when the actual learning happens. Once the algorithm obtains an initial estimate for all edge weights during the first phase, it picks up the latest index $\tau$ to start learning the optimal assignment using a synergistic blend of the well-known Upper Confidence Bound (UCB) mechanism (for MAB problems) and the Hungarian algorithm. More specifically, unlike the first phase, here, the edge weights are modified by a correction term before being fed to the Hungarian algorithm. This correction term is used to as a way to balance the tradeoff between exploration (to make up for unknown CSI) and exploitation. The term $W_{\acute{e}} = \hat{\theta}_{\acute{e}} + \sqrt{\frac{(|\acute{E}|+1)\log\tau}{\acute{m}_{\acute{e}}}}$ is a so-called confidence bound [36] and is used to inflate the estimated reward of an arm based on the level of uncertainty about the expected reward of that arm. For an arm, the inflated reward (i.e., $w_{\acute{e}}$) is called the index of that arm. Intuitively, an arm will have a high index either if it has a high estimated reward (i.e., $\theta_{\acute{e}}$) or the learner's confidence on the estimated reward of that arm is low.

| Algorithm 1 Pseudo-code of the Proposed Algorithm |
|---|
| 1: **Begin** |
| 2:     //INITIALIZATION: |
| 3:     **for** $p = 1$ $to$ $|\acute{E}|$ **do** |
| 4:         $\tau = p$; |
| 5:         Run Hungarian algorithm with arbitrary initial weights to find a link-to-cell matching $a(\tau)$ such that $p \in a(\tau)$. |
| 6:         $\left[(\hat{\theta}_{\acute{e}})_{1\times|\acute{E}|}(\hat{m}_{1\times|\acute{E}|})\right] = Evaluate(a(\tau))$; |
| 7:     **end for** |
| 8:     //MAIN_LOOP: |
| 9:     **while** 1 **do** |
| 10:         $\tau = \tau + 1$; |
| 11:         **for all** $\acute{e} \in \acute{E}$ **do** |
| 12:             $W_{\acute{e}} = \theta_{\acute{e}} + \sqrt{\frac{(|\acute{E}|+1)\log\tau}{\acute{m}_{\acute{e}}}}$; |
| 13:         **end for** |
| 14:         Run Hungarian algorithm with weights $(W_{\acute{e}})_{1\times|\acute{E}|}$ to find optimal link-to-cell matching $a(\tau)$. |
| 15:         $\left[(\theta_{\acute{e}})_{1\times|\acute{E}|}(\hat{m}_{1\times|\acute{E}|})\right] = Evaluate(a(\tau))$; |
| 16:     **end while** |
| 17: **End** |
| 18: **function** Evaluate (action $a$) |
| 19: **Begin** |
| 20:     **for** $f = 1$ $to$ $F$ **do** |
| 21:         Run the network with scheduling $a$. |
| 22:         **for** $\acute{e} \in a$ **do** |
| 23:             Record the realized $U_{\acute{e}}(x_{f,t})$. |
| 24:         **end for** |
| 25:     **end for** |
| 26:     **for** $\acute{e} \in a$ **do** |
| 27:         $U_{\acute{e}}(\tau) = \frac{1}{F}\sum_{f=1}^{F} U_{\acute{e}}(x_{f,t})$ |
| 28:         Update $(\hat{\theta}_{\acute{e}})_{1\times|\acute{E}|}$ and $(\hat{m}_{1\times|\acute{E}|})$ according to Eq. (13) and Eq. (14), respectively. |
| 29:     **end for** |
| 30: **end.** |

Exploration and exploitation are performed simultaneously by selecting the arm with the highest index at each time step. In more technical terms, the idea of this UCB-based action selection is that the square-root term is a measure of the uncertainty or variance in the estimate of an arm's value. The

---

requires either a priori knowledge of some or all statistics of the arms or hindsight information. In simulations, we use Monte Carlo to show that the regret actually does converge to zero.

quantities $w_{\acute{e}}$ (given as input to the Hungarian algorithm) are thus sort of upper bounds on the possible true values of the edges. Each time an arm is selected, the uncertainty is presumably reduced: $m_{\acute{e}}$ increments, and, as it appears in the denominator, the uncertainty term decreases. On the other hand, each time an arm other than $\acute{e}$ is selected, $\tau$ increases but $m_{\acute{e}}$ does not; because $\tau$ appears in the numerator, the uncertainty estimate increases. The use of the natural logarithm means that the increases get smaller over time, but are unbounded; all arms will eventually be selected, but the arms with lower value estimates, or those have already been selected frequently, will be selected with decreasing frequency over time.

$$\hat{\theta}_{\acute{e}}(\tau) = \begin{cases} \dfrac{\hat{\theta}_{\acute{e}}(\tau-1)m_{\acute{e}}(\tau-1)+U_{\acute{e}}(\tau)}{m_{\acute{e}}(\tau-1)+1}, & if \ \acute{e} \in \boldsymbol{a} \\ \hat{\theta}_{\acute{e}}(\tau-1), & else \end{cases} \quad (13)$$

$$m_{\acute{e}}(\tau) = \begin{cases} m_{\acute{e}}(\tau-1)+1, & if \ \acute{e} \in \boldsymbol{a} \\ m_{\acute{e}}(\tau-1), & else \end{cases} \quad (14)$$

**Remark 2:** A key assumption in our work is that the network operates in a saturated traffic condition. However, if some nodes have some inelastic (e.g., delay-sensitive) traffic, optimal scheduling needs queue-awareness in addition to channel-awareness. Given the Markovian nature of the evolution of queue lengths, handling this new setting would drastically transform the nature of the machine learning framework we have built our work on top of. In fact, in a CMAB where the edge weights are modeled by finite-state Markov chains, with unknown transition matrices, an entirely different online learning algorithm needs to be deployed to identify the optimal combinatorial structure over time. A further complication in our case would be that the Markov chains associated with the queue processes would be so-called "controlled" (i.e., affected by the decisions of the learning agent). CMAB problems with uncontrolled Markov processes have been investigated e.g., in [37]. However, we are unaware of any developments by the machine learning community for the case of CMABs with controlled processes. That being said however, our work here has taken a first step towards sparking interest in model-free optimization of TSCH-based scheduling. We have addressed CSI-free throughput maximization and this can be carried to the next level by traffic-aware delay minimization (with unknown arrival statistics).

**Remark 3:** To achieve convergence, the learning agent that acts as centralized scheduler needs to receive packet delivery feedbacks from only the links that have been scheduled during each slot-frame cycle. The volume of this information is very much less than collecting fast time-scale global CSI from across all links. In fact, while we also need a special control channel to dictate newly computed schedules and receive rate feedbacks, the amount of overhead placed over this channel is smaller than what we would have in the case of a centralized scheme which rely on instantaneous CSI, due to the following two reasons: 1) In our proposed scheme, only a subset of links are scheduled in each cycle; therefore, control information is gathered only from this particular subset. 2) The time period to dispatch the new schedules and collect feedbacks is as long as a slot-frame cycle, much longer than the period of CSI collection. Accordingly, we do not require a high capacity control medium for our algorithms, effectively cutting down on the operating expenses.

In Table 1, some of the symbols are explained.

TABLE 1
SUMMARY OF SYMBOLS

| Symbol | Description |
|---|---|
| $G$ | network graph |
| $Q$ | collision graph |
| $e \in E$ | link between two nodes |
| $n \in V$ | network nodes |
| $N$ | network size |
| $d_{i,j}$ | distance between node $i$ and $j$ |
| $w_{i,j}$ | weight of the link between node $i$ and $j$ |
| $U$ | capacity of a channel (per packet) |
| $C$ | links in collision graph |
| $\Delta$ | duration of time slots |
| $X_{f,t}$ | state of a channel $f$ in time $t$ |
| $H_{f,t}$ | gain of a channel $f$ in time $t$ |
| $B$ | bandwidth of receiving signal |
| $L$ | Packet size |
| $P(x,t)$ | power required to transmission at state $x$ in time $t$ |
| $\hat{R}_i$ | interference range of node $i$ |
| $f \in F$ | channel frequency (offset) |
| $t \in \{1, \dots, T\}$ | moment (or index) of time slots |
| $Y_t$ | signal that sent in time $t$ |
| $W_t$ | zero mean noise value |
| $\Gamma$ | number of channel different levels |
| $\mathbb{I}$ | binary decision variable |
| $R_i$ | communication range of node $i$ |
| $\tau$ | one cycle of a slot-frame |
| $\acute{e} \in \acute{E}$ | link of bipartite graph |
| $a(\tau)$ | decision vector for cycle $\tau$ |
| $X_{f,t}$ | CSI of channel $f$ in time $t$ |
| $U_{a(\tau)}$ | reward obtained during cycle $\tau$ |
| $m_{\acute{e}}$ | the number of selecting link $\acute{e}$ |
| $\hat{\theta}_{\acute{e}}$ | the estimation of weight of link $\acute{e}$ |
| $Z(\tau)$ | regret in cycle $\tau$ |
| $\acute{G}$ | weighted bipartite graph |

## V. EVALUATION

In this section, we evaluate the performance of the proposed algorithms presented in Section IV by simulating a TSCH network that includes randomly distributed sensor nodes using MATLAB. We conduct experiments with 35 nodes and compare the results.

### A. Simulation Setup

The proposed algorithms do not depend on any specific distribution for channel gain $X$, and the channel's random status is considered as an i.i.d. process. We divide the different levels of the channel into eight levels equal to the specified boundaries, see below. In fact, for each channel number and the distinct link, we consider a different probability distribution on these eight levels as assumed in [38]:
$\{(-\infty, -13 \text{ dB}), [-13 \text{ dB}, -8.47 \text{ dB}), [-8.47 \text{ dB}, -5.41 \text{ dB}),$
$[-5.41 \text{ dB}, -3.28 \text{ dB}), [-3.28 \text{ dB}, -1.59 \text{ dB}), [-1.59 \text{dB}, -0.08 \text{ dB}),$
$[-0.08 \text{ dB}, 1.42 \text{ dB}), [1.42 \text{ dB}, 3.18 \text{ dB}), [3.18 \text{ dB}, \infty)\}$
and select the channel state space as follows:
$X= \{x^1=-13 \text{ dB}, x^2=-8.47 \text{ dB}, x^3=-5.41 \text{ dB}, x^4=-3.28 \text{ dB},$
$x^5=-1.59 \text{ dB}, x^6=-0.08 \text{ dB}, x^7=1.42 \text{ dB}, x^8=3.18 \text{ dB}\}$
In Table 2, the parameters used for simulation are listed.

### 1) Evaluation Metrics

We also use the following criteria to evaluate the proposed algorithms.

- **Convergence of learning-based algorithms:** Considering the nature of learning and estimating in a learning algorithm, an important criterion is its convergence to the optimal average rate. Similarly, tending the average regret of a learning algorithm to zero can also be the criterion for the convergence study.

- **Average regret of a learning policy:** Learning policies can be evaluated on the basis of regret, and the difference between the expected reward and the reward achieved by the utilized policy. This regret is expressed in expressions (12), and here the goal is to minimize it.

- **Average throughput of each node:** In order to maximize the total throughput, according to formula (6), the average packet rate of a link can be obtained in each channel state.

- **Average total network throughput:** Using equation (15) and the average throughput of each node, the average of the total network throughput can be calculated. The term *overall_Throughput*($t$+1) is equal to the total bit rate achieved in the time slot ($t + 1$) for each link that is active in the structure of an assignment.

$$avg\_Throughput(t + 1) = \left(1 - \frac{1}{t+1}\right) avg\_Throughput(t) + \frac{1}{t+1} overall\_Throughput(t + 1) \quad (15)$$

TABLE 2
PARAMETERS AND VALUES USED IN SIMULATION

| Parameter | Value |
|---|---|
| Number of Nodes | 35 |
| Number of time-slots | 8 |
| Number of channel offsets | 3 |
| Transmission power | 10 mW |
| $WN_0$ | 2 |
| Packet size | 5000 bit |
| Cell duration ($\Delta$) | 15 mSec |

*2) Compared Approaches*

We evaluated the proposed strategies, denoted as **Perfect Statistical CSI** (or *Proposed1*) and **CSI-Free** (or *Proposed2*) by comparing their performance against three simulated baseline methods:

- **Perfect Instantaneous CSI** (*Baseline1*): In this method, we assume that in each iteration the matching process is performed using the instantaneous, complete and error-free channel information and then the best schedule in terms of the overall instantaneous throughput is calculated. Needless to add here that this is not a realistic assumption to make in real world scenarios

- **Static CSI** (*Baseline2*): In this method, assuming CSI is quasi-static, an optimal assignment between the links and the slot frame cells is calculated. In spite of random CSI changes and connection rates, the first optimal assignment for TSCH scheduling (based on obsolete CSI) will be used by the schedule ever after.

- **Erroneous CSI** (*Baseline3*): In this method, an optimal assignment of links to slot frames is computed using erroneous CSI measurements. Using a Gaussian function, each time error values are added and the optimal assignment is calculated.

It is worth noting that the assumptions made in each of the scheduling approaches in literature are very similar to those of the aforementioned baseline methods. For example, in [15 and

16], as in the first baseline method, it is assumed that an accurate instantaneous CSI is available during the scheduling process. By taking these assumptions into consideration we evaluated our proposals against these baseline methods.

*B. Simulation Results*

*1) Investigation of Convergence of Proposed Algorithm*

Figure 6 shows the average total throughput of the network obtained after one thousand repetitions for the two proposed approaches and the three baseline methods. In these experiments, the values and parameters are in accordance with Table 3. As can be seen, the second proposed method was able to converge, as in the first proposed method, within a finite number of repetition to an optimum average value.

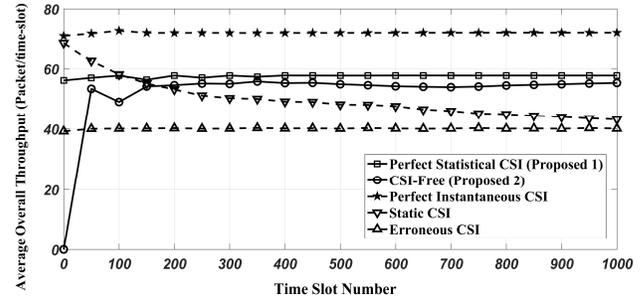

Fig. 6. Convergence Diagram of Proposed and Baseline Methods

As explained before, *Baseline1* (i.e. Perfect Instantaneous CSI) calculates the best throughput under unrealistic conditions, where perfect channel information is available. *Baseline2* (i.e. Static CSI) calculates throughput under the condition that all channel information is considered as static, in this method, throughput converges to a constant point downward. Also, *Baseline3* (i.e. Erroneous CSI), due to errors in the measurements, throughput is lower than that in other methods. For more clarity, Figure 7 shows Baseline1 individually, which expresses the oscillatory variations of the instantaneous throughput over 1000 repetitions, which finally converges to a constant value.

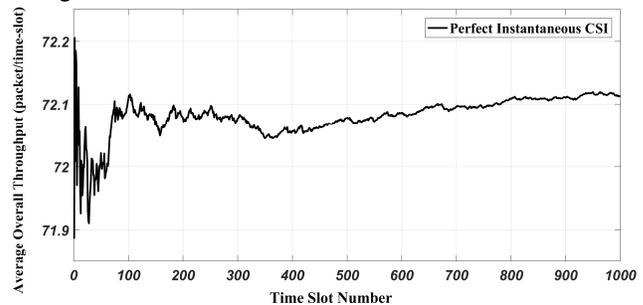

Fig. 7. Average Overall Throughput over Time

The regret of the learning algorithm in our second approach is also shown in Figure 8. As discussed in the previous chapter and is also visible in this figure, this value will tend to zero within a finite number of repetitions.

*2) Effect of Node Number on Total Network Throughput*

In this section, we evaluate scalability by increasing the number of nodes to 100 in 9 steps. Other required parameters are in accordance with Table 3. In general, the average total throughput of the network increases linearly as the number of nodes rises. Also, as can be seen in Figure 9, due to having perfect instant CSI, Baseline1 achieves highest throughput for all of the network nodes. The values of this diagram can be used as an ideal metric for evaluating performance of other methods. Proposed1 and Proposed2 are in second and third positions, respectively, in terms of average throughput. It is noteworthy that Proposed2, despite lack of access to statistical CSI, can operate similar to Proposed1 on all network metrics. Baseline methods 2 and 3 are also at the lowest efficiency levels due to lack of access to accurate CSI and unadaptive performance.

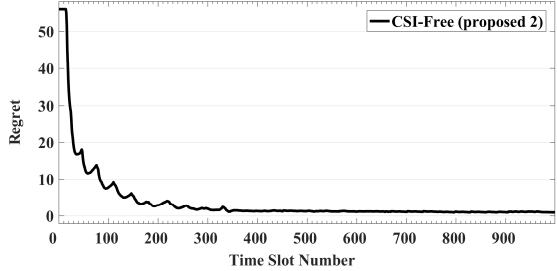

Fig. 8. (Average) Regret vs. Time

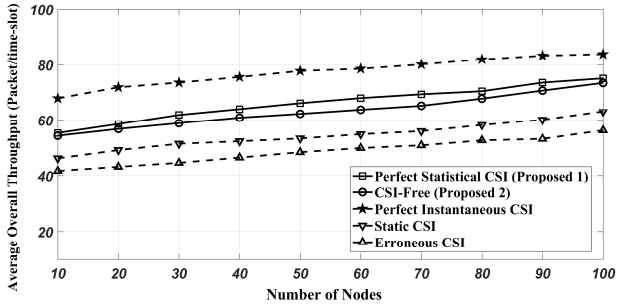

Fig. 9. Average Throughput vs. the Number of Nodes

### 3) Effect of Channel Number on Total Network Throughput

Figure 10 shows the average total throughput of the network in terms of the frequency channel offsets. In this case, 35 nodes and 8 slots are considered constant, but the number of channels varies. Other parameters are in accordance with Table 3. As can be seen, the overall trend indicates a linear increase in the average throughput of the total network as the number of channel frequency offset rises. In Baseline3, due to reliance on erroneous CSI information, we see the least amount of channel capacity utilization.

Proposed1 is more efficient compared to the ideal Baseline1 method. More importantly, Proposed2 is able to compensate for the lack of access to statistical CSI information with acceptable approximation and achieves efficiency approaching that of Proposed1.

### 4) Effect of The Slot Number on Total Network Throughput

Figure 11 shows the effect of the number of slots on the average throughput of the entire network. In this case, 10 nodes and 3 channels are considered constant, and the number of slots changes from 2 to 10. Other required parameters are given in Table 3. As can be seen, the overall effect of increasing the number of timeslots on the total network throughput is linear and upward. Similar to the other results, Baseline1 enjoys the highest average throughput in all cases. Proposed1 and Proposed2 exhibit close-to-optimal theoretical efficiency. There is a big difference between the static method, Baseline2, and the erroneous CSI-based method; Baseline3.

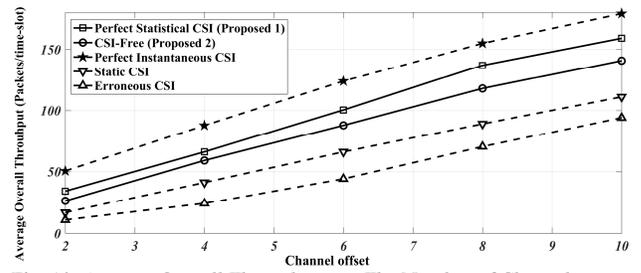

Fig. 10. Average Overall Throughput vs. The Number of Channels

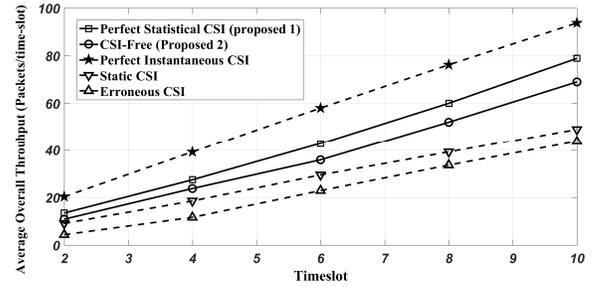

Fig. 11. Average Throughput vs. the Number of Slots

### 5) Throughput of Sender-Receiver Pairs

In Figure 12, the average throughput of each link in the two proposed solutions are evaluated. In this case 5 nodes are considered. The average throughput of 18 links is drawn as a bar graph for both methods. According to this figure, both methods are able to distribute the network capacity on the links and have managed an acceptable level of fairness.

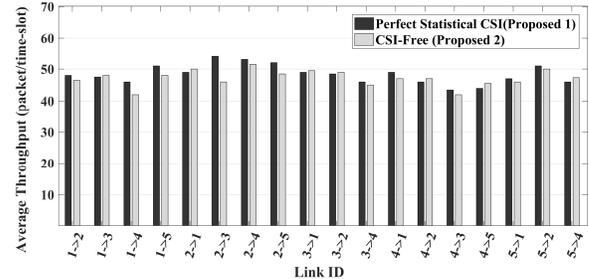

Fig. 12. Average Throughput in Different Links

## VI. CONCLUSION

We presented two new scheduling methods for TSCH networks. They did not require instantaneous CSI, and were based on statistical CSI and CSI free scheduling. In the first solution, we considered the average packet rate of connections as the weight of the bipartite graph edges. Then, using the Hungarian algorithm, we calculated the maximal weight assignment in polynomial time. In the second proposed strategy, using a learning based approach, we modeled the problem based on the CMAB framework. This strategy allowed us to apply an LLR-based solution to optimize our transmission schedules in a model free manner. These proposals do not require instantaneous information of the channel in advance, and therefore can find application in practical scenarios. Simulation results demonstrated that the second approach, which acts on the basis of feedback from previous decisions, has a short convergence time and becomes stable very rapidly. Additionally, both schemes demonstrated acceptable throughput despite lack of access to channel information, and distributed network capacity on the links in a balanced manner.

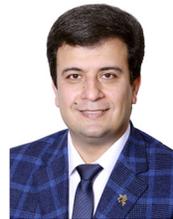

**Nastooh Taheri Javan** received his Ph.D. in computer engineering from Amirkabir University of Technology, Tehran, Iran, in 2017. Dr. Taheri Javan is currently a post-doc fellow in the Computer Engineering department at Amirkabir University of Technology. His research interests include wireless networks and network coding.

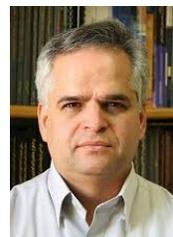

**Masoud Sabaei** is an Associate Professor with the Computer Engineering department at Amirkabir University of Technology, in Tehran, Iran. Dr. Sabaei received his Ph.D. in computer engineering from Amirkabir University of Technology, Tehran, Iran, in 2000. His research interests are wireless networks and software-defined networks.

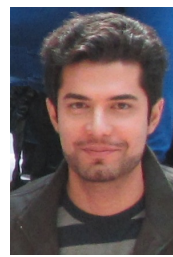

**Vesal Hakami** received his Ph.D. in computer networking from Amirkabir University of Technology, Tehran, Iran, in 2015. In 2016, Dr. Hakami joined as an Assistant Professor to the School of Computer Engineering, Iran University of Science and Technology, Tehran, Iran. His research focuses on resource control and optimization for computer networks.